\documentclass[twocolumn,12pt,cites,graphicx]{article}
\usepackage{epsfig}

\title{A basic swimmer at low Reynolds number}
\author{Marco Leoni \\
  Universit\`a degli Studi di Milano, Dip.  Fisica \\and INFN,
  sez. Milano,   Via Celoria 16, 20100 Milano, Italy \\and\\
  University of Cambridge, Cavendish Laboratory \\and Nanoscience
  Center, JJ Thomson Avenue, CB3 0HE Cambridge, UK. \and Jurij Kotar\\
  University of Cambridge, Cavendish Laboratory \\and Nanoscience
  Center, JJ Thomson Avenue, CB3 0HE Cambridge, UK. \and
  Bruno Bassetti\\
  Universit\`a degli Studi di Milano, Dip.  Fisica \\and INFN,
  sez. Milano, Via Celoria 16, 20100 Milano, Italy \and
  Pietro Cicuta\\
  University of Cambridge, Cavendish Laboratory \\and Nanoscience
  Center, JJ Thomson Avenue, CB3 0HE Cambridge, UK. \\E-mail:
  pc245@cam.ac.uk \and
  Marco Cosentino Lagomarsino\\
  Universit\`a degli Studi di Milano, Dip.  Fisica \\and INFN,
  sez. Milano, Via Celoria 16, 20100 Milano, Italy. \\E-mail:
  marco.cosentino@unimi.it} \date{
}

\begin{document}

\maketitle
\renewcommand{\thefootnote}{\fnsymbol{footnote}}

\noindent Swimming and pumping at low Reynolds numbers are subject to
the ``Scallop theorem'', which states that there is no net fluid flow
for time reversible motions. Microscale organisms such as bacteria and
cells are subject to this constraint, and so are existing and future
artificial ``nano-bots'' or microfluidic pumps. We study a very simple
mechanism to induce fluid pumping, based on the forced motion of three
colloidal beads through a cycle that breaks time-reversal
symmetry. Optical tweezers are used to vary the inter-bead
distance. This model is inspired by a theoretical swimmer proposed by
Najafi and Golestanian [Phys.Rev. E, 69, 062901, 2004], but in this
work the relative softness of the optical trapping potential
introduces a new control parameter. We show that this system is able
to generate flow in a controlled fashion, characterizing the model
experimentally and numerically.

\section{Introduction}
\label{intro}{Insight into swimming of microorganisms and bacteria can
  be gained by studying the motion at low Reynolds numbers of
  experimental and theoretical model
  systems~\cite{Ber00,Buc08}. Compared to the macroscopic world, a
  striking feature of propulsion of these micron-scale objects, for
  which inertia is typically negligible, is that a body striving to
  move has to change its shape with time in a non-reciprocal
  fashion~\cite{Pur77}.  For example, swimming of a magnetically
  driven semiflexible artificial filament was recently demonstrated,
  by a wave-like motion similar to flagella~\cite{DBR+05}.  In this
  work we realize a much simpler system which uses only two effective
  degrees of freedom, the distances between central and lateral
  spheres. This is inspired by a 3-bead linear chain theoretical
  model~\cite{NG04} which has never been realized experimentally. }

\section{Materials and Methods}
\label{materials}
\subsection{Optical tweezers}
The optical tweezers setup used in this work consists of a laser (IPG Photonics,
  PYL-1-1064-LP, $\lambda$=1064nm, P$_{max}$=1.1W) focused through a
  water immersion objective (Zeiss, Achroplan IR 63x/0.90 W), trapping
  from below. The laser beam is steered via a pair of acousto-optic
  deflectors (AA Opto-Electronic, AA.DTS.XY-250@1064nm) controlled by
  custom built electronics, allowing multiple trap generation with
  sub-nanometer position resolution. Instrument control and data
  acquisition are performed by custom software.  The trapping
  potential is locally described by a harmonic spring, and the trap
  stiffness was calibrated by measuring the thermal displacements of
  the trapped beads.  The trapping stiffness of the central bead in
  this work is $k_{\textrm{trap}}=4.1\pm1.3$ pN/$\mu$m, whereas the external
  beads were held more strongly, with $k_{\textrm{trap, \,
      external}}=8.1\pm1.4$ pN/$\mu$m. The errors quoted here are the
  standard deviation over several independent experiments, with the
  same beads and same tweezers configuration. They could also be a
  source for the slight discrepancy in relaxation times seen in
  \ref{fig: cycle}(a) and~(b). The sample is illuminated with a
  halogen lamp and is observed in bright field with a fast CMOS camera
  (Allied Vision Technologies, Marlin F-131B).  Silica beads of
  3.0$\mu$m diameter (Bangs Labs) were diluted to extremely low
  concentration to avoid any spurious bead falling into the laser
  trap. The set of three trapped beads was floated well above the
  glass slide surface (at about 10 times the bead diameter) to
  minimize any hydrodynamic drag from the solid surface. Except where
  specified differently, we used a solution of glycerol (Fisher,
  Analysis Grade) 51$\%$ by weight in water (Ultrapure grade, ELGA)
  which has a viscosity of $6.23$ mPa s~\cite{Handbook}. Experiments
  were performed at 25$^o$C.

\subsection{Experimental protocol}
  The experimental protocol is composed of two parts.  In the first
  calibration stage all the traps are kept at rest, and the beads
  undergo only Brownian motion confined by the traps. The driven
  dynamics occurs only during the second stage. Data is acquired at
  48.79~frames per second, with exposure time $19\times 10^{-3}$s. A
  run lasts ten minutes, during which we collect about 30000 frames,
  equally divided between calibration and dynamics.  We collected 4
  runs for each set of parameters.  Images are analysed using a
  correlation filter with a kernel optimized to the bead profile,
  followed by a 2-d least-square fit. This gives the center-of-mass
  coordinates of the beads in each frame with an error of the order of
  $10^{-3}\mu$m.

\subsection{Numerical simulations}
  Numerical simulations were performed by integrating directly the
  equation of motion of the three beads with Taylor's
  method~\cite{CLJ03}. We considered both simulations that included
  thermal motion (i.e. adding a white noise term to the equations) and
  deterministic ones, finding no relevant differences. The results in
  the body of the paper mainly refer to the deterministic model.

\section{Experimental}
\label{exper}We study a swimmer composed of three beads immersed in a high
viscosity Newtonian liquid. The beads are controlled by an optical
tweezer~\cite{Ash97,Blo92} and interact with each other through the
fluid.  The laser traps are set-up following the scheme of
figure~\ref{fig: intro}(a,b). The central trap is kept at rest,
holding the central bead in a soft harmonic potential of stiffness
$k_{\textrm{trap}}$, while the lateral ones switch between two positions.  By
forcing the lateral beads to move, we implement the four time-reversal
symmetry breaking phases shown in figure~\ref{fig: intro}a.

Two of the parameters that characterize the system are purely
geometric: $\epsilon$, the maximum oscillation amplitude for the
lateral beads, and $d$, the distance between the laser positions at
the starting phase of the cycle. Their physical interpretation is that
$\epsilon$ sets the strength of the stroke, while $d$ sets the
magnitude of the hydrodynamic interaction. The additional parameter is
$\tau$, the switching time of the laser, which corresponds to the
temporal length of each phase in the cycle.  Finally, the viscosity
$\eta$, which is set in the preparation of the sample, characterizes
the fluid, the hydrodynamic interactions, and the amplitude of thermal
motion.  The former parameters, together with the radius $R$ of the
beads, characterize completely the system.

Despite its simplicity, the model encapsulates all the essential
characteristics of a swimmer. We study the flow generation on the
fluid, using the central bead as a probe. This corresponds to studying
the propulsion of a freely moving swimmer.  By tracking the position
of the central bead it is possible to detect a left-right symmetry
breaking and quantify the net flow at varying parameters.

\subsection{Quantifying displacements}
Trapping lasers act with good approximation as harmonic potentials on
each bead. Thus, the average displacement $\langle \Delta x
\rangle_{t}$ of the central bead from the position of the central trap
can be converted using Hooke's law into the mean force exerted on the
fluid by the whole system.  However, a direct measurement of $\langle
\Delta x \rangle_{t}$ is at the limit of experimental resolution. The
limiting factor is not the imaging resolution, but the presence of
thermal fluctuations. This thermal noise can be compensated by longer
sampling, however to reduce this error down to the nanometer scale
would require experiments 10$^2$ times longer than ours, which is not
practical~\footnote{Already at times of the order of a few tens of
  minutes there can be sources of mechanical noise leading to
  drift.}. Instead, we quantify the flow using a different
approach. The electronically controlled trap movement temporization
allows to precisely keep track of the phase within each cycle. We can
therefore average over the repeated realizations, and reconstruct the
mean dynamic cycle of the beads. It is significant to focus on the
central one, in the stationary trap: An example of mean cycle for this
bead is shown in figure~\ref{fig: cycle}(a), where four peaks due to
the lateral motion are clearly visible.   These peaks correspond
  to the maximum displacement of the central bead in each phase of the
  motion. The configurations of the three traps are indicated
  schematically in the bottom panel of figure~\ref{fig: cycle}(a). Due
  to the different dispositions of the outer beads, the drags the
  central bead is subject during each phase are different and cause
  unequal displacements. Our method characterizes the pumping by means
  of the asymmetry of the peaks.
  On the mean cycle, we label $p_1,\, p_2,\, p_3,\, p_4$ (in order of
  decreasing value) the peaks in the displacement, and we define the
  following observables, devised to quantify the asymmetry of the
  motion: (1) $\delta_{inv}:= (p_1-p_2)-(p_4-p_3) $, the difference
  between the lower and the upper peak values, (2) $\delta_{2}:=
  p_1-p_4$, the difference between the upper and the lower maximum
  peak values.  For an illustration we refer to the mid panel in
  figure ~\ref{fig: cycle}(a).  The reason for using both definitions
  (1) and (2) to quantify asymmetry is the fact that the mean cycles
  have different shapes for different values of $\tau$. For example,
  in the case of small values of $\tau$, the intermediate peaks are
  not well distinguished, as shown in the top panel of
  figure~\ref{fig: cycle}(a). Thus, we adopt the observable
  $\delta_2$ which makes use of the upper and the lower maximum peak
  values only. In the opposite case of high values of $\tau$, see bottom panel of figure~\ref{fig: cycle}(a), the four peaks are well defined, and it
  is more effective to use the observable $\delta_{inv}$. The
  advantage is that $\delta_{inv}$ is determined independently from
  the equilibrium position $x_0$, which is needed for $\delta_{2}$ and
  cannot be measured during the active motion.  Thus, it allows for
  more direct measurements.  These observables lack a simple physical
  meaning and should rather be thought as order parameters useful in
  characterizing the asymmetry of the peaks. Both these quantities
  change sign by reverting the sequence of steps and allow to verify
  if the system behaves as expected, pumping in the opposite
  direction. We also define a third observable (3) $s:= |p_1| + |p_2|
  + |p_3| + |p_4|$, the sum of absolute values of the peaks.  $s$ is
  always positive, and quantifies the amplitude of the swimmer's
  motion.

\subsection{Comparison of experiments with simulations}
We compare the measured flow to simple numerical simulations (see
methods). The hydrodynamic interaction between the spheres is
described by the Oseen tensor~\cite{DE86}, corresponding to the limit
of point force, or far field. Along the $\hat{x}$ axis of the swimmer,
the equation of motion for bead $n$ is
\begin{eqnarray}
\label{eq:motion eq}
\dot{x}_n &=& \frac{1}{\gamma} \left(  F_n + \sum_{n \neq m}
  \left(\frac{3 R}{2 r_{nm}}\right)  F_m \right)
\quad  \nonumber \\  &\textrm{with}& \quad n,m=1,2,3,
\end{eqnarray}
where $\gamma = 6 \pi \eta R$ is Stokes' drag, $F_i$ is the external
force exerted on the $i$-th particle, and $r_{nm}$ indicates the
relative distance between the $n$ and $m$. The experimental values of
$\eta$ and $k_{\textrm{trap}}$ were used in the simulations.


As can be seen by inspecting figure \ref{fig: cycle}(b), the observed
relaxation times  are shorter in the experimental data of bead motions
than in the simulations at corresponding parameters.  To exclude
possible errors, we performed the following tests.  First, we
simulated a single particle in a stationary trap, including thermal
fluctuations, calculated the relaxation time in the time autocorrelation
function. For an isolated particle, the relaxation time is the ratio
${k_{\textrm{trap}}}/{\gamma}$. Using the experimental values in the
simulations, we found excellent agreement for beads both in pure water
and water/glycerol.
Subsequently, we compared the simulation (with and without thermal
noise) to the experiment for one trapped particle in water/glycerol, in a
trap undergoing repeated displacements, see figure 2(c). Also in this
case, the simulation and experiment match very well. In conclusion,
there is excellent agreement in the single particle regime, confirming
experimental calibration and numerical methods.  Therefore the slight
discrepancy in the three-bead data must be due to the excessively
simple theoretical description of multiple interacting bodies%
%
~\footnote{Note that we can exclude that inertia and the propagation
  time of hydrodynamic interaction plays a role, as deviations were
  found to be at the nano-second scale~\cite{HMB02,AKS+05}, well
  outside from the time scales of our experiments.}.
%
%
In other words, the flow induced by each bead is not perfectly
represented by a linear superposition of single-particle Oseen
propagators~\cite{DE86}.  This is not surprising, considering that
this is a long-distance, or far-field approximation.  A common
procedure to correct for this effect makes use of a perturbative
expansion in the parameter ${R}/{d}$. We checked this by implementing
the first and second perturbative corrections~\cite{RP69,Bat82} in our
simulations, finding very similar quantitative results as with the
Oseen tensor.  The reason for this is that the parameter ${R}/{d}$ is
not small, being of the order of ${1}/{3}$, so that the real dynamics
is not accessible perturbatively~\footnote{While approaches that do
  not make use of the explicit form of the solution of hydrodynamic
  equations~\cite{MB74,LM05} may help solving this problem, in
  general, to compensate for this small discrepancy one would have to
  solve Stokes' equation with the proper boundary
  conditions~\cite{ADL07}.}.
Despite this limit, the simulations do allow us to compare the
experimental results with a prediction for the net force, or flow,
generated in the fluid by the swimmer.  A mapping between each of
  the observables ($\delta_2$, $\delta_{inv}$) and the temporal
  average $\langle \Delta x \rangle_t$ is obtained by means of
  simulations. Here, sampling problems are not present and $\langle
  \Delta x \rangle_t$ is quantifiable directly by time averaging with
  arbitrary accuracy; $\delta_2$ and $\delta_{inv}$ are obtained from
  the steady-state mean cycle, similarly to experiments.  For both
  $\delta_2$ and $\delta_{inv}$, we verify that there exists a one to
  one mapping with $\langle \Delta x \rangle_t$.  We use this relation
  to associate as a function of $\epsilon$ each point of the
  experimental curve to the corresponding mean displacement, obtaining
  $\langle \Delta x \rangle_t$ as a function of $\epsilon$ that is
  converted into a mean force via Hooke's law.  The procedure is
  illustrated in Figure~S1 of supplementary methods.

\section{Discussion}
Our first result is that the system is able to generate flow.  This is
qualitatively visible in the distribution of position of the central
bead (figure~\ref{fig: intro}(c)). This effective potential is
composed of two contributions: the harmonic potential exerted by the
tweezer, and the configurational bias induced by the interaction with
the lateral moving beads. The latter contribution causes the emergence
of two effective asymmetric minima. This asymmetry can be quantified
accurately using the above-defined observables (figure~\ref{fig:
  results}).  Simulations show that there exists a one-to-one map
between $\delta_{inv}$ and $\delta_{2}$, and the values for the mean
force, so that it is possible to convert $\delta_{inv}$ and
$\delta_{2}$ directly to forces.  For fixed values of $\tau$, we ran a
series of experiments to characterize the dependence of the mean
position from the displacements $\epsilon$ of the lateral beads.
Figure \ref{fig: intro}(b) shows one of the two possible time-reversal
breaking sequences.  By reverting the sequence into the specular one,
the swimmer should move in the opposite direction.  Experimentally, we
implemented both sequences, and found that this property holds.
Figure \ref{fig: results}(a) shows a comparison between the
experimental and simulated observable $\delta_{inv}$ as a function of
$\epsilon$.  The absolute value of the corresponding mean force is
shown in figure~\ref{fig: results}(b).
The maximum mean forces reached in our experiments are of the order of
0.03pN, roughly corresponding to a net swimming/pumping speed of about
0.2$\mu$m/s. Note that this force is exceedingly small to be measured with
an optical tweezer using conventional techniques, which has posed a
barrier in past investigations~\cite{RWG+97}. The indirect technique
used here enables to overcome this barrier, and can possibly be useful
in different contexts.

It is interesting to study the influence of $\tau$, by fixing the
value of $\epsilon$, and in figure \ref{fig: results}(c) we plot the
 propulsion force at varying $\tau$, compared with simulations. In
both cases there is a good quantitative agreement between the model
with Oseen interactions and the experimental data. By increasing the
distance $d$ between the beads, the asymmetry in the peaks becomes
more difficult to detect.
However it is still possible to characterize the dependence of fluid
pumping on the parameter $d$ by using the amplitude variable $s$. In
figure~\ref{fig: results}(d), $s$ is plotted versus the distance, for
both simulations and data.  In both cases this quantity decays as
$1/d$. Quantitatively, there is a small systematic deviation between
experiment and simulations. The $1/d$ decay directly depends on
hydrodynamic interactions, and can be understood with a simple
argument on the maximum deviation of an initially resting trapped bead
subject to the perturbation induced through the fluid by another bead
at distance $d$ relaxing for a stretch $\epsilon$ in another harmonic
potential (see Supplementary Material for further details).

\subsection{Comparison with other model systems}
 It is instructive to compare our swimmer
with the closely related theoretical model originally proposed in the
literature~\cite{NG04} by Najafi and Golestanian, and the variants
that have been recently explored~\cite{GA08,Gol08}.  There are two
main features characterizing our model.
Firstly, beads are subject to the compliance of the trapping
potentials. The original Najafi-Golestanian swimmer has rigid
links. More recently, the model has been solved for the case of
chemical transitions between states~\cite{GA08} and for general bead
and link sizes and imposed deformations~\cite{ADL07, Gol08}.  However,
an actuated movement by quadratic potentials has not been addressed
explicitly in the previous literature.  While the behavior of the
simplest rigid Najafi-Golestanian swimmer is essentially characterized
by the ratio $\epsilon/d$~\cite{NG04,Yeo07}, more general models can have
more complex dynamics, imposed by additional relevant time or
length-scales. Our system is also subject to an additional parameter,
the ratio $\frac{\tau}{\tau_0}$, with $\tau_0 := \gamma/k_{\textrm{trap}}$ of
the imposed displacement time to the characteristic relaxation time of
the bead. Hence the phenomenology of our system depends also on the
ratio of the two time scales.
The most important consequence is that our simulations show a
power-law dependence of the speed, or mean force, on $\epsilon$, with
an exponent that increases monotonically with increasing $\tau$,
slightly larger than the quadratic law found for the Golestanian
swimmer. Experimentally, it is problematic to measure incontrovertibly
this scaling, due to the large errors in the fit for the exponents,
whose average values, however, do exceed two.
In the limiting case of small $\tau$, the two models behave in the
same way. This can be understood in the following way. Assume for
simplicity that the lateral forcing beads can be described as free
particles relaxing trough the minimum of an harmonic potential. Their
position follow a simple temporal law governed by an exponential decay
$x(t) \sim x_0 \,\exp(-{t}/{\tau_0})$, and accordingly the velocity
$v(t) \sim ({x_0}/{\tau_0}) \exp(-{t}/{\tau_0})$.  For small values of
$\tau$, the ratio ${\tau}/{\tau_0}$ is small ensuring that the
velocity of the trapped bead does not decay significantly during each
step in the cycle, being a constant proportional to $\tau_0$ as in the
Najafi Golestanian swimmer, where the beads have a constant
displacement velocity~\cite{NG04}.
On the other hand, our results confirm that in the small $\tau$ limit
the mean force scaling with $\epsilon/d $ is compatible with a
quadratic law.  This regime is also the one where the swimmer's
propulsion is optimal.  In
the numerical experiments, the fitted exponents we find are consistently larger
than 2 (between 2.12 and 2.41).  However, the predicted trend with
$\epsilon$ cannot be confirmed experimentally because of the large errors  (Supplementary Figure~S2).
The second difference with the original Najafi-Golestanian swimmer is
that our swimmer is not intrinsic, but actuated by external
forces~\cite{DBR+05,LCL03}. In particular, the total actuating force
is not null instantaneously, but only over a cycle.  This has been
shown to lead to qualitatively distinct behavior in other
systems~\cite{Lau07}. In our case, using simulations of an analogous
intrinsic swimmer actuated by two-body two-state springs, we find that
the differences are mostly quantitative, the intrinsic variant being
slightly faster and more efficient (Supplementary Figure~S3).

Finally, it may be of interest to compare this pump to other
engines. We can estimate the dissipation rate in our pump to be
between $0.5\times10^{-17}$ and $4\times10^{-17}$J/s, which is
comparable to an \textit{E.~coli} bacterium swimming in
water~\cite{Bus05}. The efficiency of this pump is of the order of
$10^{-3}$\% (calculated as the ratio of the mean output to input
power) which is lower than for typical biological systems such as
\textit{E.~coli}~\cite{Ber00,Wu06}. These results are important on
fundamental grounds and for applications in nano-scale machines.\\

\textbf{Acknowledgments}. We thank Ray Goldstein, Christopher Lowe and Ignacio Pagonabarraga for
useful comments and discussion. This work was supported by the Korea Foundation for International
Cooperation of Science \& Technology (KICOS) through a grant provided
by the Korean MOST (No.2007-00338) and by the UK EPSRC.

\clearpage

\onecolumn

\begin{figure}[t]
\includegraphics[width=8.5cm]{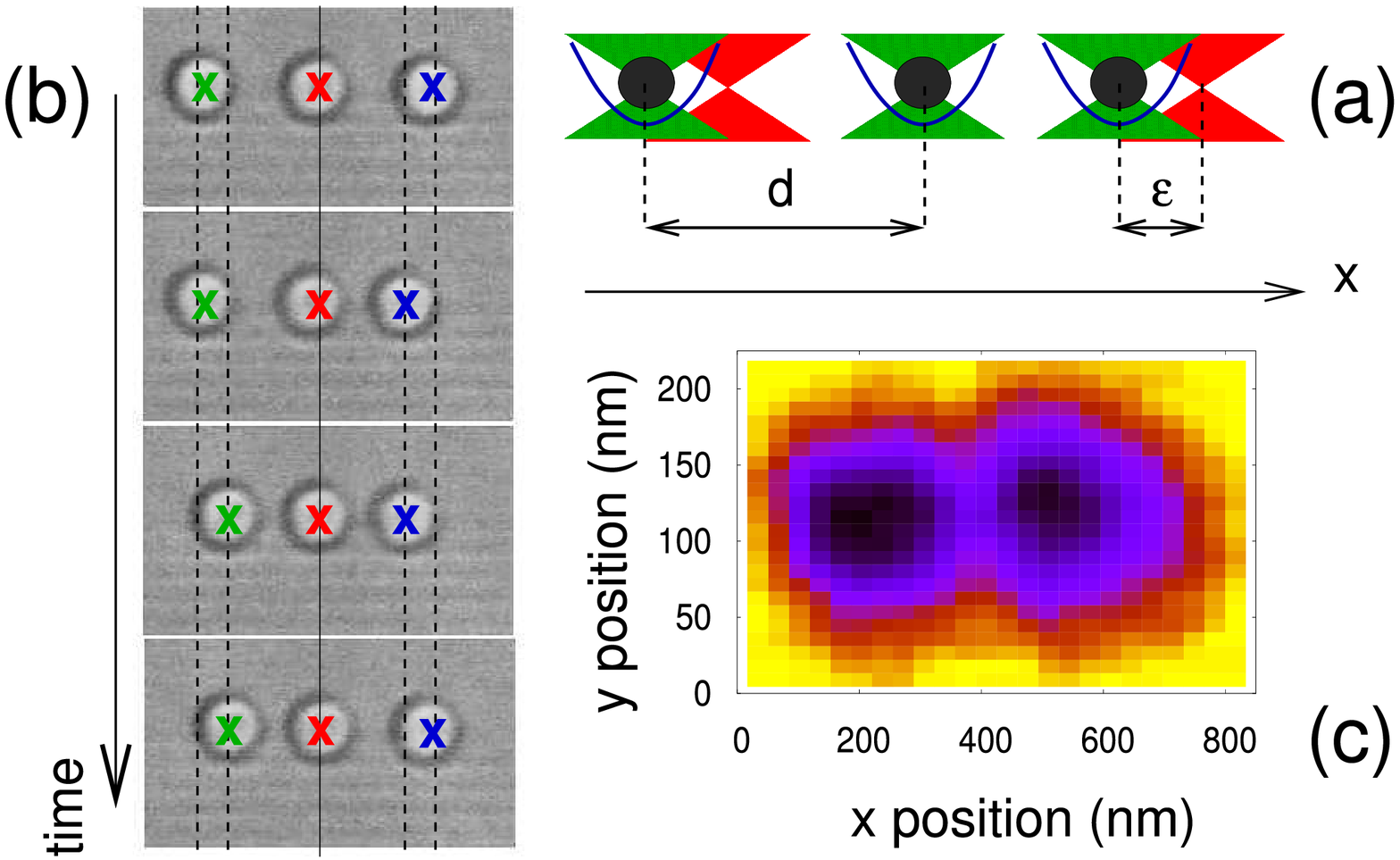}
\caption{{\bf Experimental realization of the low Reynolds number
    swimmer, and proof of force generation.}
 \label{fig: intro}}
\end{figure}

\begin{figure}[t]
\includegraphics[width=8.5cm]{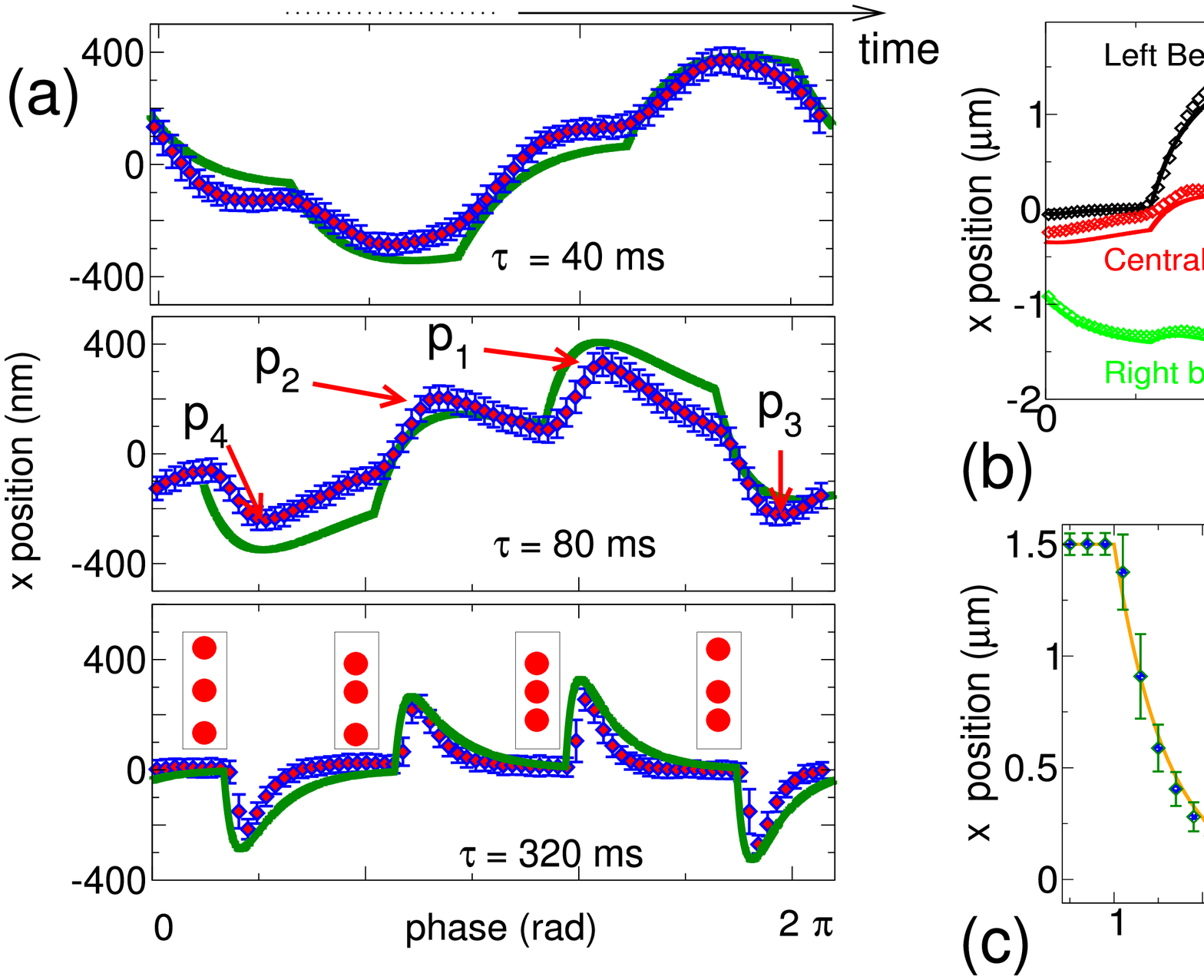}
\caption{{\bf Reconstruction of the mean cycle.}   \label{fig: cycle} }
\end{figure}

\begin{figure}[t]
\includegraphics[width=8.5cm]{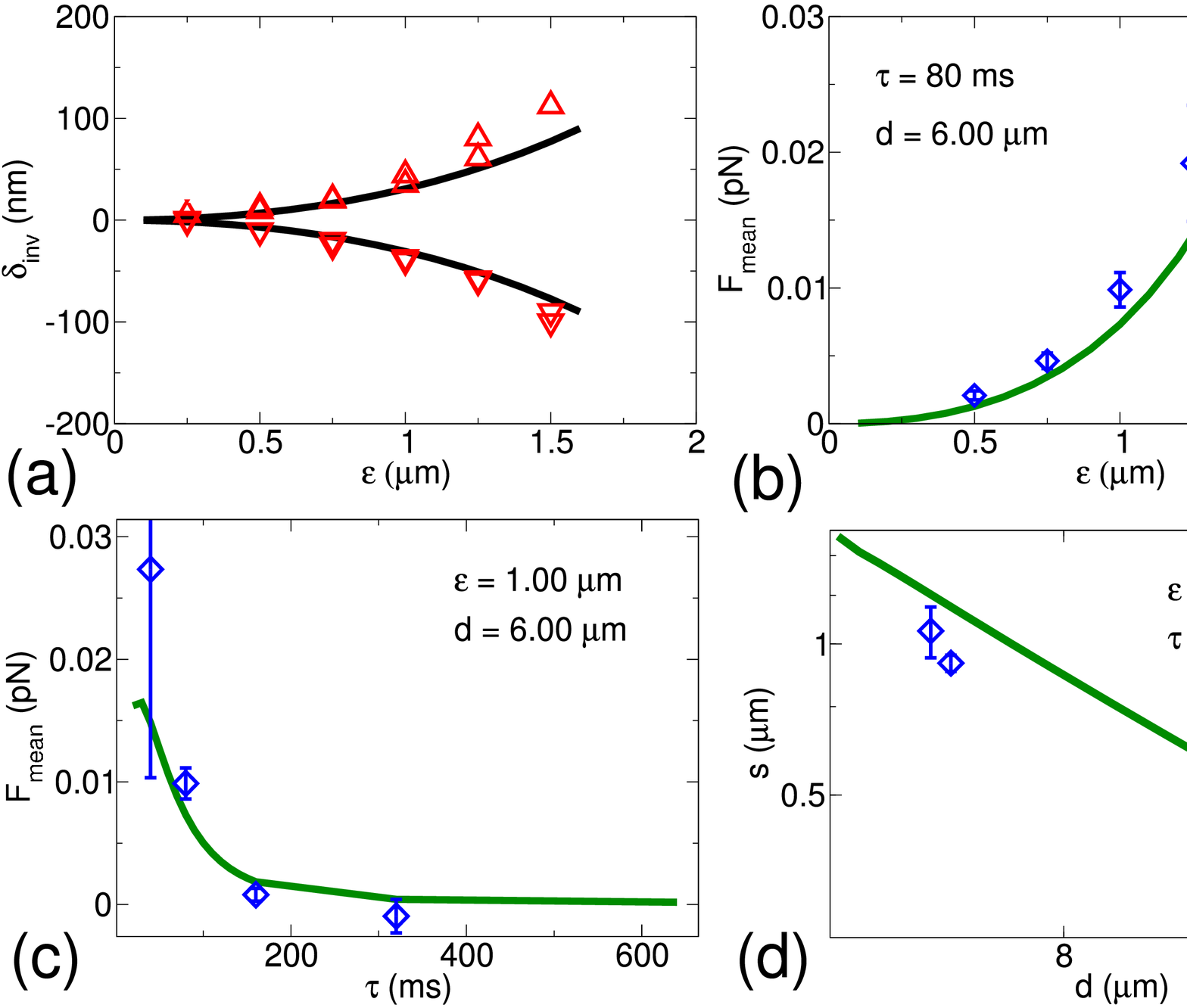}
\caption{{\bf Quantitative characterisation of the swimmer.} \label{fig: results}}
\end{figure}

\begin{list}{}{\leftmargin 2cm \labelwidth 1.5cm \labelsep 0.5cm}

\item[\bf Fig. 1] {\bf Experimental realization of the low Reynolds number
    swimmer, and proof of force generation.}
  (a) Scheme of the lasers' disposition during a single phase of
  motion, in which the active traps (green) act on the beads as
  harmonic potentials. The parameters $\epsilon$ and $d$ are sketched.
  (b) Sequence of snapshots of the experiment showing four
  time-symmetry breaking steps in the basic cycle. Crosses indicate
  the active laser trap positions.  By moving the lateral beads and
  keeping the central one at rest, it is possible to study the flow
  generated on the fluid, using the central bead as a probe. A
  left-right asymmetry in its position signals that flow is generated.
  Top panel: starting configuration, where the distances
  between beads take their maximum values.
  (c) Two-dimensional map of the effective potential felt by the
  central bead. The plot is obtained by taking the logarithm of the
  relative frequencies of the positions occupied by the central bead
  during 500 repetitions of the basic cycle.  The colour scale
  indicates in dark the most frequent position. The map shows a
  left-right symmetry breaking (the left-hand minimum is deeper),
  which is a proof of the mean generated flow in the system.

\item[\bf Fig. 2] {\bf Reconstruction of the mean cycle.} The
  experiment is a sequence of repetitions of the elementary cycle
  illustrated in figure~\ref{fig: intro}.  By treating each period as
  a single realization and averaging over all realizations, one
  recovers the mean cycle of the dynamics, averaged over the thermal
  fluctuations due to Brownian motion in the fluid.
  (a) Detail of the mean cycle for the central bead ($\diamondsuit$
  with errorbars) compared with simulations (solid lines) for
  different $\tau$. The plot reports the longitudinal displacement
  coordinate of each bead as a function of the cycle phase in the
  interval $(0, 2 \pi)$. The peaks become more defined as $\tau$
  increases. Top panel: $\tau=40$ ms ; middle panel: $\tau=80$ ms ;
  bottom panel: $\tau=320$ ms. The position of the four peaks and the
  traps' configuration in the four phases of the cycle are illustrated
  in the middle and bottom panel respectively.
  (b) Mean cycle for the whole system (three beads). The longitudinal
  displacement coordinate of each bead is plotted as a function
  of time.  Dotted lines correspond to experimental values for the
  left bead (black), the central bead (red) and the right bead
  (green). These are compared with the results of numerical simulations (solid lines with the same color code).
  (c) Close up and complete cycle showing relaxation of a single bead
  in a switching trap, with no interacting partners.

\item[\bf Fig. 3] {\bf Quantitative characterisation of the swimmer.}
  The main observable is the mean force felt by the central bead,
  measured by its mean displacement in the optical trap. The mean
  force is studied here as a function of the parameters $\epsilon$ and
  $\tau$. (a) shows $\delta_{inv}$ as a function of the
  lateral amplitude $\epsilon$, comparing experimental values
  (triangles) with simulation (continuous line). There are two
  possible time-reversal-symmetry breaking cycles, corresponding to
  the two opposite pumping directions. By temporally reversing the
  cycle showed in figure~(1), $\delta_{inv}$ changes sign,
  showing that the direction of pumping can be controlled. The
  following panels show experimental data ($\diamondsuit$) compared to
  simulations (continuous line). Each $\diamondsuit$ represents the
  average over 4 different experiments and the error bars correspond
  to the standard deviations of the four values. (b)~Shows the mean
  force (obtained from the observable $\delta_{inv}$ and
  numerical simulations) as a function of $\epsilon$, for $d=6 \mu$m,
  $\tau=80$ ms, $\eta=6.23$ mPa\,s.  (c)~Mean force as a function of
  $\tau$.  Here the other experimental parameters are $d=6 \mu$m,
  $\epsilon=1 \mu$m, $\eta=6.23$ mPa\,s.  (d) Amplitude of motion as a
  function of $d$. Increasing $d$ the force becomes weaker and
  difficult to detect, and we have to use the amplitude parameter $s$
  to compare with simulations. The measured and simulated values of
  $s$ scale as $d^{-1}$ (as discussed further in the Supplementary Material).

\end{list}

\clearpage


\clearpage
\newpage
\section*{\centering \Large SUPPLEMENTARY MATERIAL}
\section*{Additional considerations on the model and the data analysis}

\renewcommand{\thesection}{S\arabic{section}}
\setcounter{figure}{0}
\setcounter{table}{0}
\setcounter{section}{0}
\renewcommand{\figurename}{Supplementary Figure}
\renewcommand{\thefigure}{S\arabic{figure}}
\renewcommand{\tablename}{Supplementary Table}
\renewcommand{\thetable}{S\arabic{table}}

\vspace{4cm}
\section*{Scaling with $d$}

We present here a simple argument for the scaling of the amplitude $s$
with distance $d$, explaining  the scaling seen for experiments and simulations in figure~3(d).

To estimate the $d$-dependency of the amplitude during a cycle, it is
sufficient to describe the displacement of the central bead in one of
the four sub-phases. We consider the simplified but physically
equivalent situation of two interacting trapped beads at distance $d$.
Initially the left bead, which represents the central bead in the
experiment, is at rest and the right bead is out of its equilibrium
position by a distance $\epsilon$, as it would be if in the instant
the trap has been shifted. The initial conditions are thus $x_L(0)=0,
x_R(0)=d$ and the equilibrium positions of the trapping potentials are
$x_{0,L}=0; x_{0,R}=(d-\epsilon)$. To a first approximation, (``zeroth
order'' in $R/d$) the left bead is still, and the position of the
right bead follows a simple relaxation law
\begin{equation}
  x_R^{(0)}(t) = \epsilon \left( e^{-\frac{t}{\tau_0}} -1 \right) + d
  \ \ .
\end{equation}
This unperturbed solution can be used to estimate (by the force
balance with the fluid) the source of force $F_R^{(0)} = -
k_{\textrm{trap}} (x_R^{(0)} - d) $ applied by the right bead on the
fluid during its relaxation. This can then be used into equation~(1) of the main text for the left bead, giving
\begin{equation}
  \tau_0 \ \dot{x}^{(1)}_L = x^{(1)}_L -\frac{3 R}{2 d}
    x_R^{(0)} \ \ ,
\end{equation}
where $ x_R^{(0)}$ appears as an external perturbation, and we
approximated the distance between the beads with $d$, which is
justified in the limit of large distances $d >> \epsilon$.  With this
assumption, the problem becomes linear and the solution to order $R/d$
can be calculated as
\begin{equation}
  {x}^{(1)}_L(t) = \frac{3 R \epsilon}{2 d}
  \frac{t}{\tau_0} e^{-\frac{t}{\tau_0}}  \ \ .
\end{equation}
In turn, this solution could be used as a source for the equation for
$x_R$, to obtain hierarchically the higher order contribution in $R/d$
to its motion.  The value of the peak of the central bead in the
experiment $s$ can be estimated by the maximum displacement of the
left bead
\begin{equation}
  x^{(1)}_{L,\mathsf{max}} = \frac{3R \epsilon}{2d e} \ \ .
\end{equation}
This argument implies that the leading order scaling of each peak in a
cycle, and hence of $s$, is $1/d$. The argument has the advantage of
showing how the hydrodynamic interaction tensor comes into play
explicitly after a trap switches its position.\\In the
same linear approximation, the coupled equations (eq.1) from the main text
can even be solved directly in a straightforward way, and the
resulting maximum displacement is:
\begin{equation}
  x_{L,\mathsf{max}} = \frac{3R \epsilon}{2d} \frac{1}{\left(1+
      \frac{3R}{2d}\right)^{1+ \frac{2d}{3R}} }\ \ ,
\end{equation}
which has the same behavior in the limit of large $d/R$.

Strictly speaking, this result is applicable in the regime $\tau >>
\tau_0$, in which the beads have the time to fully relax in the trap
potentials. In the opposite limit, $\tau << \tau_0$, since the
subcycle ends while the central bead is still moving away from the
center of its trap, the maximum position can be estimated using the
same solution, ${x}^{(1)}_L(t)$, by the position assumed by the
central bead at the end of this subcycle, i.e. the instant
$t=\tau$. Thus as
\begin{equation}
  x^{\prime(1)}_{L,\mathsf{max}} =  \frac{3R \epsilon}{2d} \frac{\tau}{\tau_0}
  e^{-\frac{\tau}{\tau_0}} \ ,
\end{equation}
and the scaling with $d$ is unaffected.

\clearpage
\newpage
\section*{Relating the observables $\delta_2$ and $\delta{inv}$ with
  the mean force.}
  
  We summarise here how the procedure to relate the experimental data to the induced flow.

\begin{figure}[h!]
\centering
\includegraphics[width=0.6\textwidth]{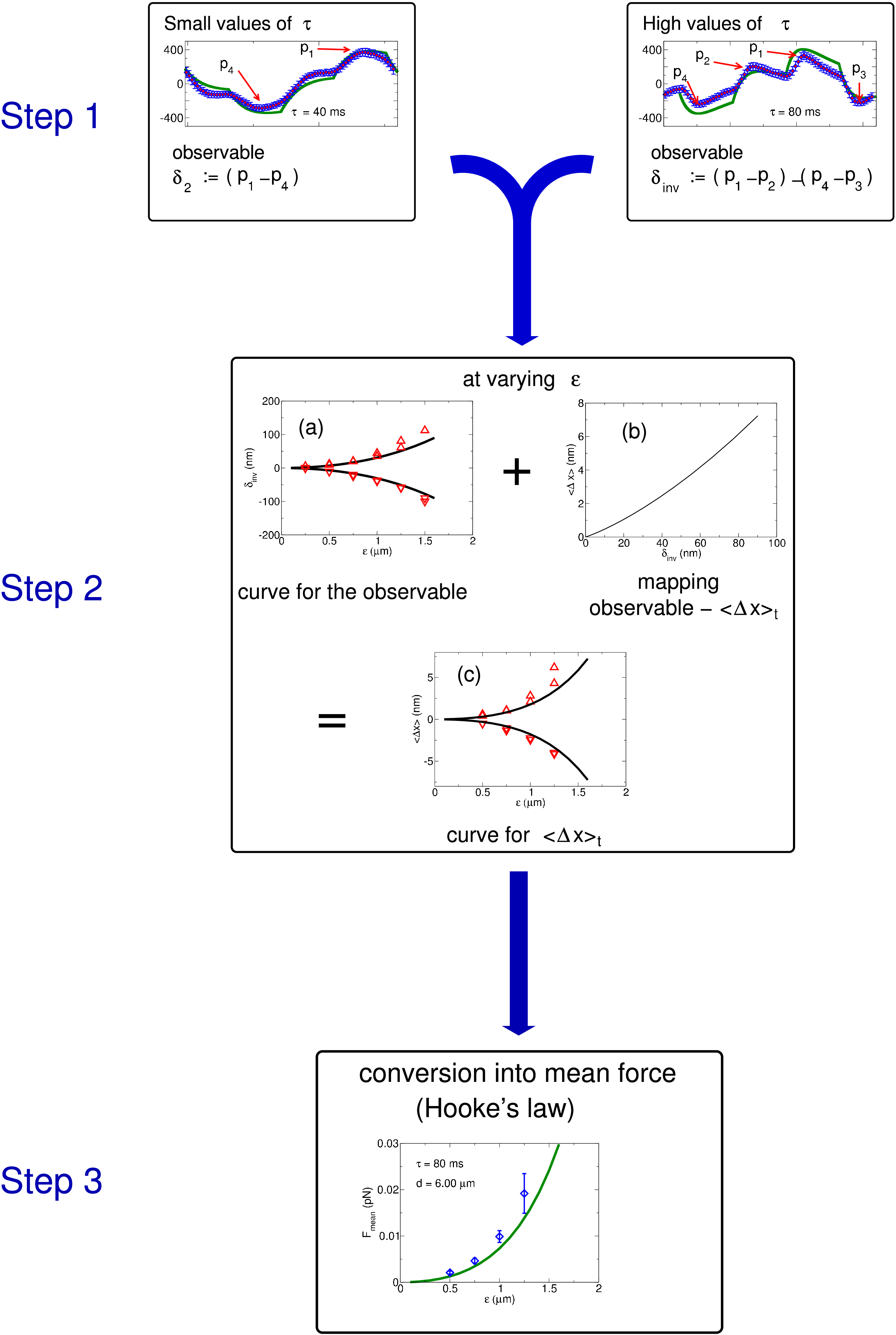}
\caption{{\bf Converting observables into mean force.} This scheme
  illustrates the procedure used to obtain the mean force from the
  experiments, by means of simulations. Step 1: Because of the
  different shapes of the mean cycles for different values of $\tau$,
  we define two different observables $\delta_2$ and $\delta{inv}$ to
  quantify the asymmetry in the displacements for the central
  bead. The peaks which enter in the definitions are indicated with
  red arrows. Step 2: (a) We analyze the mean cycle of experiments and
  simulations, extracting each observable at varying $\epsilon$. We
  also calculate directly the temporal average $\langle \Delta x
  \rangle$ for the position of the central bead. (b) Comparing these
  results and eliminating the dependence from $\epsilon$, we find that
  there exist a one to one mapping between each observable and the
  temporal average $\langle \Delta x \rangle$. (c) Using this relation
  the curve of each observable as a function of $\epsilon$ can be
  converted into a curve of the mean position as a function of
  $\epsilon$. Step 3: using Hooke's law we convert the mean
  displacement into a mean force as a function of $\epsilon$.}
\label{fig:our_method}
\end{figure}
\clearpage
\newpage

\section*{Scaling with $\tau$}
\begin{figure}[tbh!]
\centering
\includegraphics[width=0.6\textwidth]{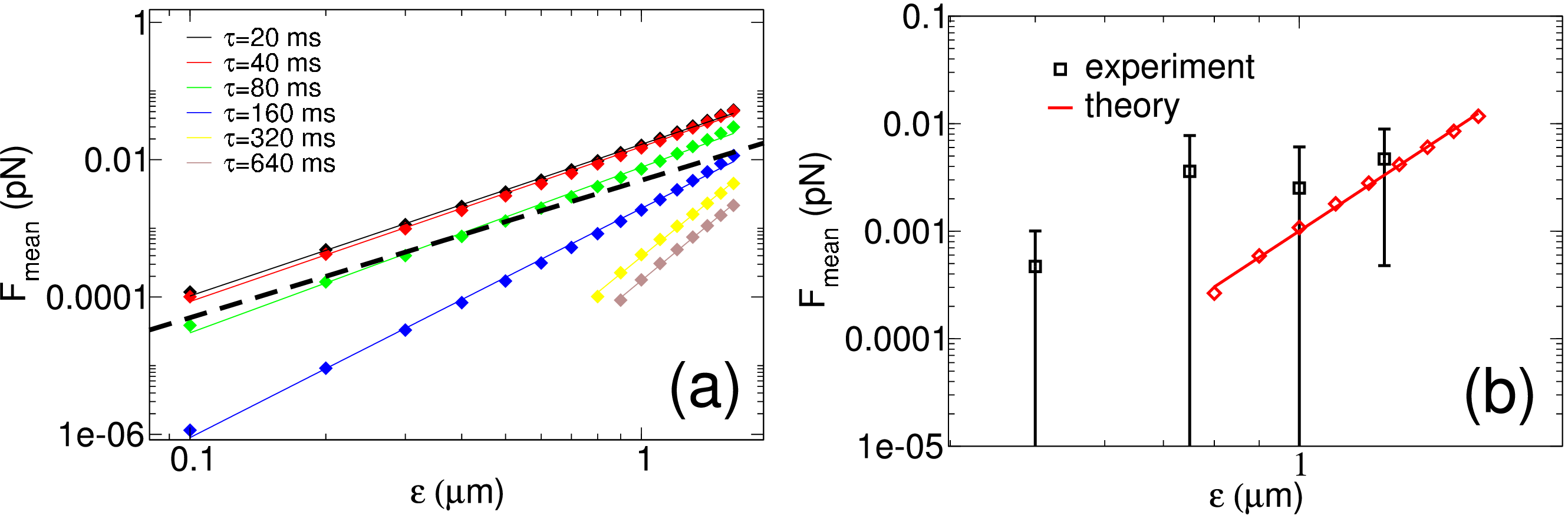}
\caption{{\bf Scaling law for the mean force at varying $\tau$}. (a)
  Plot on log-log scale of the simulated mean displacement from the
  equilibrium position of the central bead. The curve is a power law
  with different exponents for different values of $\tau$. In our model the
  exponent varies, and increases monotonically for increasing
  $\tau$. In the limit of small $\tau$ the mean displacement follows a
  power law with exponent close to 2 (dashed line), which resembles the behavior of
  the Golestanian swimmer. (b) Comparison between experimental and
  theoretical data for $\tau = 320$ ms on log-log scale. Due to the
  large error bars on the experimental curve, the determination of the
  exponent from the experiment is subject to large uncertainties. \label{fig:scaling}}
\end{figure}
\clearpage
\newpage

\section*{Intrinsic Swimmer}
\begin{figure}[h!]
\centering
\includegraphics[width=0.6\textwidth]{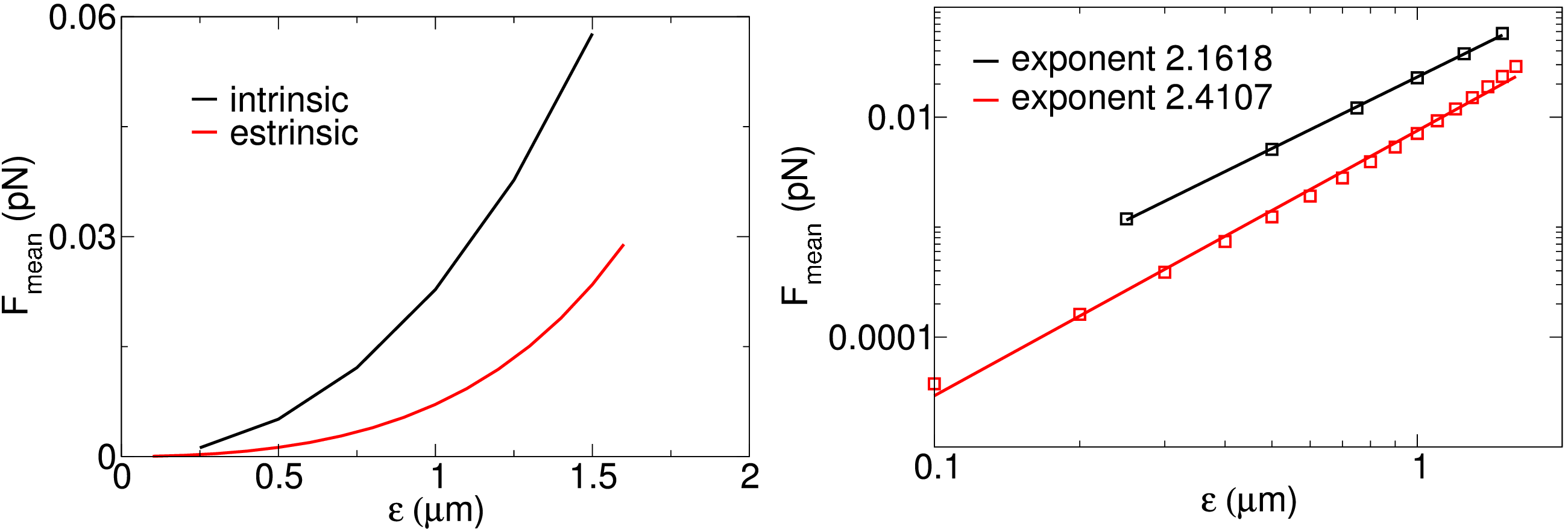}
\caption{{\bf Comparison of the propulsive forces for extrinsic and
    intrinsic swimmers.} The extrinsic swimmer is studied
  experimentally and numerically in this work, whereas we can only
  study the intrinsic swimmer, actuated by two-state springs,
  numerically. The plot shows the mean force with varying $\epsilon$,
  for simulations of the two models, with parameters $d=6\mu$m
  $\tau=80$ms. The difference between the two swimmers is only
  quantitative. \label{fig:intri_estri}}
\end{figure}

Figure \ref{fig:intri_estri} shows a comparison of our model with
simulations of an analogous intrinsic swimmer.

\end{document}